\documentclass[11pt,a4paper]{article}

\usepackage[utf8]{inputenc}
\usepackage[margin=2.5cm]{geometry}
\usepackage{amsmath,amssymb}
\usepackage{graphicx}
\usepackage{bm}
\usepackage{xcolor}
\usepackage{authblk}
\usepackage[super,sort&compress,comma]{natbib}
\usepackage[labelfont=bf,labelsep=period]{caption}
\usepackage{hyperref}
\hypersetup{
    colorlinks=true,
    linkcolor=blue,
    citecolor=blue,
    urlcolor=blue
}

\newcommand{\SiN}{\mathrm{SiN}_x}

\newcommand{\Xzero}[1]{X^{0}_{#1}}
\newcommand{\Xminus}[1]{X^{-}_{#1}}
\newcommand{\AP}{\mathrm{AP}}
\newcommand{\RP}{\mathrm{RP}}
\newcommand{\dEAR}{\Delta E_{\AP\text{-}\RP}}

\renewcommand{\figurename}{Fig.}

\title{Optical characterization of Fermi polarons in monolayer MoSe$_2$ encapsulated in amorphous silicon nitride}

\author[1,2,*]{Christos Paspalides}
\author[3]{Brendan C. Mulkerin}
\author[4]{Rafael Luque Merino}
\author[1]{Alvaro Castilla Lamas}
\author[1]{Samuel Palacios Valdepe\~nas}
\author[5]{Kenji Watanabe}
\author[6]{Takashi Taniguchi}
\author[3]{Jesper Levinsen}
\author[3]{Meera M. Parish}
\author[1,2,7]{Luis Vi\~na}
\author[1,2,7]{Maria Dolores Mart\'in}
\author[8,2,7,*]{Francesca Maria Marchetti}
\author[4]{Andres Castellanos-Gomez}
\author[1,2,7,*]{Carlos Ant\'on-Solanas}

\affil[1]{Departamento de F\'isica de Materiales, Universidad Aut\'onoma de Madrid, 28049 Madrid, Spain}
\affil[2]{Instituto Nicol\'as Cabrera, Universidad Aut\'onoma de Madrid, 28049 Madrid, Spain}
\affil[3]{School of Physics and Astronomy, Monash University, Clayton, Victoria 3800, Australia}
\affil[4]{2D Foundry Research Group, Instituto de Ciencia de Materiales de Madrid (ICMM-CSIC), 28049 Madrid, Spain}
\affil[5]{Research Center for Electronic and Optical Materials, National Institute for Materials Science, 1-1 Namiki, Tsukuba 305-0044, Japan}
\affil[6]{Research Center for Materials Nanoarchitectonics, National Institute for Materials Science, 1-1 Namiki, Tsukuba 305-0044, Japan}
\affil[7]{Instituto de F\'isica de la Materia Condensada, Universidad Aut\'onoma de Madrid, 28049 Madrid, Spain}
\affil[8]{Departamento de F\'isica Te\'orica de la Materia Condensada, Universidad Aut\'onoma de Madrid, 28049 Madrid, Spain}
\affil[*]{Corresponding authors: christos.paspalides@uam.es, francesca.marchetti@uam.es, carlos.anton@uam.es}

\date{}

\begin{document}

\maketitle

\begin{abstract}
Amorphous silicon nitride is the leading low-loss, CMOS-compatible platform for wafer-scale integrated photonics. Encapsulating transition-metal dichalcogenide (TMD) monolayers within this platform offers a promising route towards scalable TMD-photonic circuits. Yet the effect of this amorphous dielectric on the optical response of the monolayer remains poorly understood. Here, we compare monolayer MoSe$_2$ encapsulated in amorphous silicon nitride with a hexagonal boron nitride-encapsulated reference, using differential reflectivity and photoluminescence. Whereas the reference is dominated by the neutral exciton, the nitride-encapsulated monolayer shows attractive and repulsive polaron resonances separated by ${\sim}31$~meV, with emission dominated by the attractive branch. A quantitative comparison of temperature-dependent reflectance with a microscopic finite-temperature Fermi-polaron theory shows that the nitride induces substantial electron doping in the monolayer,
with an electron density of $(0.8$--$0.9)\times10^{12}$~cm$^{-2}$. Using the extracted density and trion binding energy as the only input parameters, the theory reproduces the photoluminescence spectra up to ${\sim}80$ K in excellent quantitative agreement with experiments,
thus corroborating the Fermi-polaron description of the observed excitonic response.
\end{abstract}

\section{Introduction}

The integration of transition metal dichalcogenide (TMD) monolayers with photonic circuits is a promising route towards compact and scalable light sources, modulators, and nonlinear elements. Silicon nitride is the natural platform for this goal: it is the workhorse dielectric of low-loss, CMOS-compatible integrated photonics, combining wafer-scale deposition, foundry availability, and tight optical confinement in waveguides and resonators, with broadband transparency spanning the exciton transitions of most TMDs~\cite{blumenthal2018,bi2026}. Routing and multiplexing have been demonstrated on this platform~\cite{yang2014,xiao2024}. Monolayer TMDs have accordingly been interfaced with silicon nitride in many geometries of interest: waveguides and resonators, for modulation, nonlinear conversion, and emission control~\cite{datta2020,joshi2020,parto2022,rarick2024,kumarmandal2023,huang2025}; and patterned membranes and metasurfaces, for strong light-matter coupling and photoluminescence (PL) enhancement~\cite{cao2020,deng2026}. The silicon nitride, as a photonic platform, may also be engineered to host its own intrinsic single-photon emitters~\cite{Farheen2026}. In these directions, hexagonal boron nitride (hBN) offers a less versatile route: the growth and transfer of large-area, single-crystal films have yet to reach the maturity required for seamless integration with photonic-foundry processing~\cite{mckeowngreen2024,wang2026wafer}.

In nearly all of these demonstrations, however, the monolayer sits in direct contact with the nitride, and its optical response has not been fully characterized. This is important because the Coulomb interaction in a two-dimensional crystal is only weakly screened, and thus the properties of the excitonic and charged-exciton complexes are highly sensitive to the dielectric and electrostatic environment of the monolayer~\cite{mak2010,wang2018,chernikov2014}. This sensitivity can be a resource, since it allows the response to be engineered, but it can also become a liability, since disorder at the monolayer-substrate interface---in the form of surface roughness, physisorbed residues, and substrate charge traps---inhomogeneously broadens the resonances~\cite{rhodes2019}, obscuring information about the intrinsic properties of the monolayer. hBN is the standard remedy: its atomic flatness, chemical inertness, and wide band gap allow encapsulated monolayers to reach PL and absorption linewidths approaching the homogeneous limit, with residual charge disorder essentially eliminated~\cite{dean2010,cadiz2017,ajayi2017}. A recent comparative study of monolayer WS$_2$ found substantially greater inhomogeneous broadening in silicon nitride environments than in hBN, attributing the difference primarily to substrate-induced charge trapping~\cite{shao2022}. Whether silicon nitride can serve as an effective protective and encapsulating layer for TMDs, using hBN as the benchmark, therefore hinges on its electrostatic influence on the monolayer.

In particular, sub-stoichiometric amorphous nitride (SiN$_x$) grown by plasma-enhanced chemical vapor deposition has been shown to induce substantial $n$-type doping in adjacent WSe$_2$ flakes~\cite{chen2014}, and dielectric-induced charge doping has recently been reported for MoS$_2$ on hBN, ZrO$_2$, and SiO$_2$ substrates~\cite{Loh2026}. More recently, studies of MoSe$_2$ at a GaInP interface~\cite{shan2026} have shown that charged excitonic complexes acquire substantial oscillator strength in absorption and exhibit suppressed carrier recoil in PL, consistent with their identification as attractive Fermi-polaron resonances. Other studies have demonstrated substrate-induced doping using flexible conducting redox-polymer substrates and strain-tunable plasmonic metasurfaces for controlling and probing Fermi-polaron resonances~\cite{Wu_ACSNano_2026}. However, the doping induced by SiN$_x$ encapsulation remains largely unexplored, with previous estimates relying on indirect indicators such as linewidth broadening~\cite{shao2022}.

In this work, we compare monolayer MoSe$_2$ encapsulated in hBN and amorphous SiN$_x$ using differential-reflectivity and PL spectroscopy, and we establish that SiN$_x$ encapsulation induces substantial electron doping in the monolayer. In particular, we observe that the SiN$_x$-encapsulated monolayer exhibits both attractive (AP, trion-like) and repulsive (RP, exciton-like) Fermi-polaron resonances, whereas the optical response of the hBN-encapsulated reference is dominated by the neutral exciton. Through a quantitative comparison with a microscopic finite-temperature Fermi-polaron theory~\cite{tiene2023,mulkerin2023}, we estimate the electron density and trion binding energy, which are then used to describe the PL spectra at different temperatures with no additional fitting parameters. The theory reproduces the AP emission profile well at low temperatures, where the RP emission is very weak, and yields excellent agreement with both the AP and RP resonances at the highest temperatures. These results provide direct optical evidence for substrate-induced charge-transfer doping in SiN$_x$-encapsulated TMDs, and quantify its influence on the excitonic response of monolayer MoSe$_2$. Since the optical dielectric constants of the two hBN and SiN$_x$ encapsulants are similar, the results further show that it is the charged defects of the amorphous SiN$_x$ network that set the excitonic response of the monolayer.

\section{Results}

\subsection{hBN versus SiN$_x$ encapsulation}

\begin{figure}[t]
\centering
\includegraphics[width=0.62\textwidth]{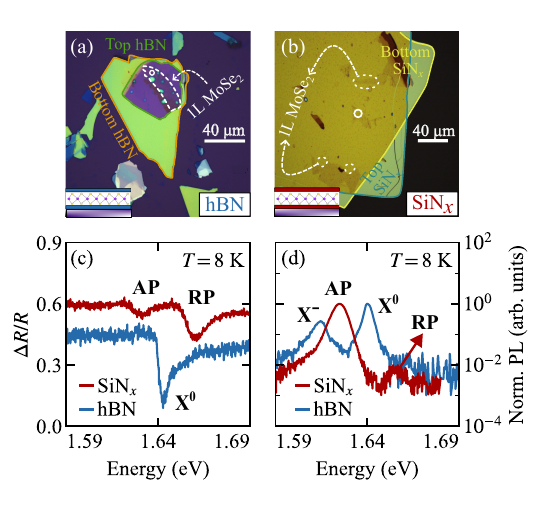}
    \caption{\textbf{hBN- versus SiN$_x$-encapsulated monolayer MoSe$_2$ at $T=8$~K.} \textbf{a}~hBN/1L~MoSe$_2$/hBN reference sample (top and bottom hBN flakes outlined in green and orange, respectively, and monolayer in dashed white).
    \textbf{b}~SiN$_x$/1L~MoSe$_2$/SiN$_x$ sample (top and bottom membranes illustrated by cyan and yellow solid lines respectively, monolayer regions dashed). In both \textbf{b} and \textbf{c}, the white circle within the monolayer region indicates the location at which the reflectance and PL measurements were performed, while the inset depicts the cross section of the sample.
    \textbf{c}~Differential reflectivity, $\Delta R/R$, of the SiN$_x$ (red) and hBN (blue) samples, where an offset between the traces is added for the sake of clarity. \textbf{d}~PL spectra of the same samples, each normalized to its maximum, on a logarithmic scale. $\Xzero{}$ and $\Xminus{}$ denote neutral and trion excitons in the sample with hBN capping, respectively; AP and RP denote attractive and repulsive polarons in the SiN$_x$-encapsulated monolayer MoSe$_2$ sample.}
\label{fig:reference_samples}
\end{figure}

Figure~\ref{fig:reference_samples} compares the two reference samples: MoSe$_2$ monolayer encapsulated in hBN (Fig.~\ref{fig:reference_samples}a) and in SiN$_x$ (Fig.~\ref{fig:reference_samples}b). The absorptive response, measured as differential reflectivity at $T=8$~K, is shown in Fig.~\ref{fig:reference_samples}c. The hBN-encapsulated monolayer displays a single, well-isolated resonance at $1.6425$~eV, which we assign to the neutral exciton $\Xzero{}$; no charged-exciton feature is resolved above the noise, indicating a low density of resident carriers. The SiN$_x$-encapsulated monolayer instead exhibits two $\Delta R/R$ resonances: a stronger one at $1.657$~eV and a weaker one ${\sim}31$~meV below it, which we identify as the RP and AP polaron branches, respectively. The sizeable AP oscillator strength provides spectroscopic evidence of a nonzero resident electron density, since the AP acquires oscillator strength only in the presence of carriers~\cite{sidler2017,efimkin2017}.

The corresponding PL spectra are shown in Fig.~\ref{fig:reference_samples}d. For the hBN-encapsulated monolayer, PL is dominated by $\Xzero{}$ with the trion $\Xminus{}$ appearing ${\sim}25$~meV below it at roughly a quarter of the peak intensity, the standard signature of a weakly and unintentionally doped MoSe$_2$ monolayer~\cite{ross2013,mak2013}. The emission in the SiN$_x$-capped monolayer is instead dominated by the AP: a single line at $1.6235$~eV, with the RP barely visible in log scale at ${\sim}1.655$~eV. Furthermore, the AP-RP splitting
is larger than the exciton--trion splitting in the hBN-capped sample,
consistent with a larger electron density~\cite{sidler2017,efimkin2017}.
The AP dominance of the PL emission is not restricted to the measured spot: spatially resolved PL maps show the AP line as the dominant emission across several disconnected monolayer islands of the same stack (Supplementary Note~2 and Supplementary Fig.~2).

\subsection{Carrier density from temperature-dependent reflectance}

\begin{figure}[t]
\centering
\includegraphics[width=0.62\textwidth]{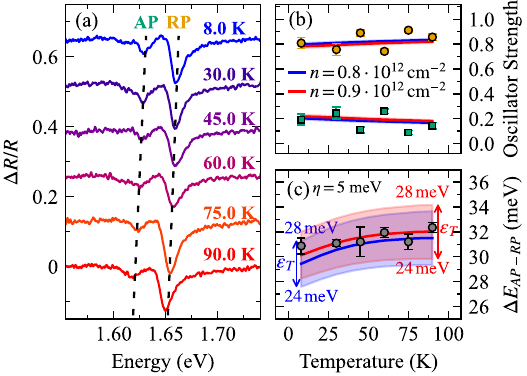}
    \caption{\textbf{Temperature dependence of reflectance spectra of the SiN$_x$-encapsulated monolayer and comparison with the Fermi-polaron theory.} \textbf{a}~$\Delta R/R$ spectra for different temperatures, from $T=8$~K to $90$~K, where a trace offset is introduced for the sake of clarity. Two dashed lines indicate the AP/RP resonances as a guide to the eye. \textbf{b}~Normalized oscillator strengths $Z_{\AP}$ (green symbols) and $Z_{\RP}$ (orange symbols) extracted from the reflectance spectra; solid lines are the calculated $Z_{\AP,\RP}$ for $n=0.8$ (blue) and $0.9$ (red) $\times10^{12}$~cm$^{-2}$. \textbf{c}~AP-RP energy splitting $\dEAR$: experiment (symbols) and theory for the same two densities, with the same colour code as in \textbf{b}; shaded areas span $\varepsilon_T=24$--$28$~meV and solid lines correspond to the central value, $\varepsilon_T=26$~meV. Error bars in \textbf{b} and \textbf{c} are obtained from a Monte Carlo analysis of the spectral fits (Methods).}
\label{fig:abs_theory}
\end{figure}

To quantify the doping, we investigate the temperature evolution of the reflectance of the SiN$_x$-encapsulated monolayer from $8$ to $90$~K (Fig.~\ref{fig:abs_theory}a). Both resonances are resolved at all temperatures and redshift in parallel by ${\sim}10$~meV, following the temperature dependence of the band gap of MoSe$_2$~\cite{vina1984} (Supplementary Fig.~4). From each spectrum we extract the AP--RP splitting, $\dEAR$, and the oscillator strengths $Z_{\AP,\RP}$, defined as the integrated spectral weights of the two resonances normalized such that $Z_{\AP}{+}Z_{\RP}=1$; error bars are obtained from a Monte Carlo analysis (Methods). We find that the splitting changes little, from $30.9\pm0.6$~meV at $8$~K to $32.4\pm0.4$~meV at $90$~K (Fig.~\ref{fig:abs_theory}c),
while the measured AP oscillator strength fluctuates between $Z_{\AP}=0.09$ and $0.26$
without a strong systematic dependence on temperature (Fig.~\ref{fig:abs_theory}b).

We apply the same procedure to reflectance spectra calculated with the Fermi-polaron model (Methods), which depend on two unknown parameters, the electron density $n$ and the trion binding energy $\varepsilon_T$. No calculation of $\varepsilon_T$ is available for MoSe$_2$ encapsulated in SiN$_x$. Since the optical dielectric constant of the nitride is close to that of hBN (Supplementary Note~5), we explore the interval $\varepsilon_T=24$--$28$~meV, around the $X^0$--$X^-$ separation of our hBN reference.
Fitting the temperature dependence of the experimental oscillator strengths $Z_{AP,RP}$ (Fig.~\ref{fig:abs_theory}b) and energy splitting $\Delta E_{AP-RP}$ (Fig.~\ref{fig:abs_theory}c) to the theoretical predictions, we obtain $n=(0.8$--$0.9)\times10^{12}$~cm$^{-2}$ and $\varepsilon_T\simeq26$~meV. This substantial electron doping is comparable to that achieved with other doping substrates~\cite{shan2026,Wu_ACSNano_2026}, while the
trion binding energy is
close to the $X^0$--$X^-$ separation of the hBN-encapsulated  reference, ${\approx}25$~meV, as expected from the similar dielectric screening of the two environments. A loss of electrons from the monolayer with increasing temperature, e.g., by thermal emission into traps of the nitride, would transfer oscillator strength from the AP to the RP and reduce the splitting. Neither occurs: a single, temperature-independent density accounts for both observables (splitting and relative area) up to $90$~K.

The doping has a natural origin in sub-stoichiometric SiN$_x$ that is grown by plasma-enhanced chemical vapor deposition.
In this case, it was shown that positively charged $^{+}\mathrm{Si}\!\equiv\!\mathrm{N}_3$ dangling-bond centers can transfer an electron density of up to $1.75\times10^{12}$~cm$^{-2}$
to an adjacent WSe$_2$ monolayer~\cite{chen2014}, which is consistent with the extracted densities here.
Dielectric-induced doping has also been reported for MoS$_2$~\cite{Loh2026}, and charge trapping by the nitride was identified as the main source of inhomogeneous broadening in SiN$_x$-supported WS$_2$~\cite{shao2022}.

\subsection{Photoluminescence and the polaron-to-trion crossover}

\begin{figure}[t]
\centering
\includegraphics[width=0.55\textwidth]{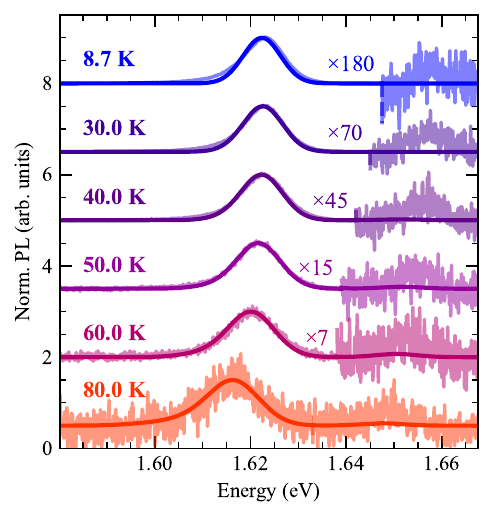}
    \caption{\textbf{Temperature dependence of the PL for SiN$_x$-encapsulated monolayer and comparison with Fermi-polaron theory.} Measured PL spectra (light colors) for different temperatures, from $8.7$ to $80$~K. Solid dark lines: PL calculated with the Fermi-polaron theory using the parameters obtained from reflectance, $n=0.9\times10^{12}$~cm$^{-2}$ and $\varepsilon_T=26$~meV.
    Curves are normalized to their maxima and vertically offset for clarity. The weak emission corresponding to the RP at ${\sim}1.65$~eV is magnified  (for both theory and measured curves) as indicated, except for $T=80$~K.}
\label{fig:PL_theory}
\end{figure}

Turning to PL, we see in Fig.~\ref{fig:PL_theory} that the emission at low temperatures is dominated by the AP, while the RP progressively gains emission intensity as the temperature increases.
This agrees with the expectations based on theory, since the emission spectrum is related to the exciton spectral function $A(\omega)$ through the so-called detailed-balance relation $P(\omega)\propto e^{-\hbar\omega/k_BT}A(\omega)$~\cite{kennard1918} (see Method section).
In other words, the relative intensities of the two branches are determined by their oscillator strengths, weighted by the corresponding thermal occupation factors. We find that the magnification required to resolve the RP emission decreases from $\times180$ at $8.7$~K to $\times7$ at $60$~K, while at $80$~K it is directly resolved.
Over the same range, the AP line broadens from $10$ to $22$~meV FWHM, mostly above $50$~K, and its integrated intensity falls by a factor of ${\sim}60$ (Supplementary Fig.~4). The low-temperature linewidth is the same over the whole scanned monolayers and independent of the explored excitation powers (Supplementary Figs.~2 and~3).

With the parameters extracted from reflectance, the theory reproduces the PL spectra in the entire range from $8.7$ to $80$~K (Fig.~\ref{fig:PL_theory}): namely, the relative energy between AP and RP peaks, the width of the AP line, its thermal broadening, and the emergence of the weak RP emission from $40$~K upwards. Within the explored range of doping density $n$ and trion binding energy $\varepsilon_T$, the calculated spectra differ mainly in the weak RP branch, which is best seen on a logarithmic scale (Supplementary Note~7): a larger $\varepsilon_T$ shifts the RP further from the AP and lowers its intensity. The PL model therefore supports the parameter set obtained from reflectance.
At $8.7$ and $30$~K, small discrepancies between theory and experiment arise in the low-energy tail of the AP emission, which is broader in the experimental spectra. This broadening can be attributed to a low-energy disorder tail induced by the $\SiN$ encapsulation, which is not accounted for in the model.
Furthermore, the discrepancy between theory and experiment in the RP emission intensity at low temperatures may arise from a slow  thermalization rate between different polarons such that the populations have not fully equilibrated before the exciton decays.

Extracting both the doping density and the trion binding energy allows us to compare the relevant energy scales of the problem, namely the Fermi energy $E_F$, $\varepsilon_T$ and $k_BT$, and thus to identify the physical regime probed by our experiments.
Finite-temperature Fermi-polaron theory predicts that, for  $E_F\lesssim\varepsilon_T$, the nature of the AP is controlled by the ratio $E_F/k_BT$~\cite{tiene2023}. In the quantum-degenerate regime, $E_F\gg k_BT$, the AP is a well-defined quasiparticle. When $E_F \lesssim 0.7\,k_BT$, however, the trion--hole continuum subsumes the attractive branch such that the AP loses its quasiparticle character.
This temperature-driven polaron-to-trion crossover implies a qualitative change in the shape of the AP PL emission profile, even though the AP--RP splitting and oscillator strengths remain relatively constant~\cite{tiene2023}.
In our SiN$_x$-encapsulated monolayer, taking $m_e=0.6\,m_0$, the extracted density corresponds to $E_F=\hbar^2\pi n/m_e\approx3.2$--$3.6$~meV,
and the crossover condition
is reached at $T\approx55$--$60$~K. Therefore, the attractive branch evolves from a polaron quasiparticle towards the trion-continuum dominated regime across the temperature series of Figs.~\ref{fig:abs_theory} and~\ref{fig:PL_theory}.
Indeed, both the thermal broadening of the PL line reproduced in Fig.~\ref{fig:PL_theory}
and the observed onset of the AP broadening above $50$~K (Supplementary Fig.~4) are direct signatures of the polaron-to-trion crossover.

\section{Concluding remarks}

To conclude, we have found that encapsulating monolayer MoSe2 in amorphous SiN$_x$ drives it into a doped Fermi-polaron regime. A Fermi-polaron theory with a temperature-independent electron density, $n=(0.8$--$0.9){\times}10^{12}$~cm$^{-2}$, accounts for the distribution of oscillator strength, the AP--RP splitting, and the PL spectra up to $80$~K, demonstrating that the charge transferred from the nitride persists over this range.
Importantly, doping from charged defects of the amorphous nitride is expected to depend on the stoichiometry and the deposition conditions, which must therefore be controlled when integrating TMD monolayers on silicon-nitride photonic circuits.
Combined with SiN$_x$ resonators and metasurfaces that already reach exciton-photon strong coupling~\cite{cao2020}, the Fermi sea provided by the nitride offers a gate-free route to Fermi polaron-polaritons on a platform compatible with wafer-scale integrated photonics.

\section{Methods}

\subsection{Sample fabrication}

Both heterostructures were assembled by deterministic transfer techniques. The SiN$_x$ membranes are commercial, $50$-nm-thick, ultra-low-stress amorphous silicon nitride windows (PELCO, Ted Pella), with a specified surface roughness of $0.65$~nm~\cite{tedpella}. They were released from their Si frames with the nail-polish technique~\cite{haley2021} and placed on the target substrate. The hBN crystals were grown by the high-pressure, high-temperature method~\cite{taniguchi2007}, and the MoSe$_2$ crystals were purchased from HQ Graphene. MoSe$_2$ and hBN flakes were mechanically exfoliated with adhesive tape onto PDMS stamps and dry-transferred at room temperature~\cite{CastellanosGomez2014}. Both samples were then annealed in vacuum at $200\,^\circ$C for $2$~h. Throughout this work, SiN$_x$ denotes this amorphous, chemical-vapor-deposited nitride, from which our membranes are made, as opposed to crystalline Si$_3$N$_4$. We note that, unlike atomically flat hBN, the membranes leave roughness as a possible source of inhomogeneous broadening~\cite{shao2022}. The supplier describes the films as non-stoichiometric but considers their deposition method and composition proprietary, and has no data on their refractive index or fixed-charge density; we treat these properties as unknown (Supplementary Note~5).

\subsection{Optical spectroscopy}

Optical measurements were performed in an open-loop helium cryostat (CryoVac) with the sample held at temperatures between $8$ and $90$~K. Spectra were recorded with a SpectraPro 2500i spectrometer (Acton Research) coupled to a PIXIS $1024\times1024$ CCD. PL was collected through a $20\times$, ${\rm NA}{=}0.45$ objective (Olympus LCPlan N IR) following off-resonant excitation with a green, continuous-wave laser ($532$~nm); residual laser light was rejected with a $550$~nm long-pass filter. For differential reflectivity, the sample was illuminated with a white lamp and $\Delta R/R = (R_{\rm ML}{-}R_{\rm ref})/R_{\rm ref}$ was obtained from spectra taken on ($R_{\rm ML}$) and off ($R_{\rm ref}$) the monolayer area. Spatially resolved maps were acquired by scanning the excitation spot across a fixed collection path using a $4f$ relay and a steering mirror, which images the mirror onto the back focal plane of the objective~\cite{Lee2007}, with the PL signal collected by a $20\times$, ${\rm NA}{=}0.40$ objective (Mitutoyo M Plan Apo NIR).
For momentum-resolved polarization tomography, a flip-mounted lens images the back focal plane of the objective onto the spectrometer slit, and the emission is analysed by motorized half- and quarter-wave plates followed by a fixed linear polarizer; the normalized Stokes parameters are reconstructed pixel by pixel from six polarization projections of the spectrally integrated AP line. The polarization properties of the AP emission are further discussed in Supplementary Note~6. Full schematics of both optical configurations are given in Supplementary Fig.~1.

\subsection{Spectral analysis}

The oscillator strengths $Z_{\AP,\RP}$ were obtained as the integrated areas of the two $\Delta R/R$ resonances above a local baseline, normalized to unit sum, which removes the common optical prefactor of the stack, and $\dEAR$ as the separation of the two reflectivity minima. The uncertainties of $\dEAR$ and $Z_{\AP,\RP}$ reflect the choice of baseline and of the integration limits of the two resonances. They were estimated by a Monte Carlo procedure in which these choices are varied randomly within reasonable bounds and the analysis is repeated; each error bar is the standard deviation of the resulting distribution. The same procedure was applied to the calculated reflectance spectra. Details of the analysis of the PL maps and of the azimuthal analysis of the Stokes parameters are given in Supplementary Notes~2 and~6.

\subsection{Fermi-polaron theory}

We model the optical response of an electron-doped MoSe$_2$ monolayer using the finite-temperature Fermi-polaron framework developed in Ref.~\citenum{tiene2023}, extending earlier zero-temperature theories of exciton polarons~\cite{efimkin2017,sidler2017}. In this framework, the optically excited excitons are dressed by particle-hole excitations of the thermal Fermi sea of excess electrons, leading to AP and RP resonances whose energies and optical spectral weights are controlled by the electron density $n$,
the temperature, and the trion binding energy. The optical response is determined by the exciton Green's function, $G_X(\omega)=[\hbar\omega-\Sigma(\omega)+i0^+]^{-1}$, where the energy $\hbar\omega$ is measured relative to the bare exciton resonance. The self energy $\Sigma(\omega)$ encodes the many-body dressing of the exciton by the surrounding electron gas, and is given by $\Sigma(\omega)=\sum_{\mathbf{q}} f_{\mathbf{q}} T (\mathbf{q} , \hbar\omega+\epsilon_{\mathbf{q}}^{e})$ in terms of the exciton-electron $T$ matrix,
\begin{equation}
T^{-1} (\mathbf{q} , \hbar\omega) =- \sum_{\mathbf{k}} \frac{1}{\varepsilon_T + \epsilon_{\mathbf{k}}^{e} + \epsilon_{\mathbf{k}}^{X}} - \sum_{\mathbf{k}} \frac{1-f_{\mathbf{k}}}{\hbar\omega - \epsilon_{\mathbf{k}}^{e} - \epsilon_{\mathbf{k} - \mathbf{q}}^{X} + i0^+}.
\end{equation}
Here, $\epsilon_{\mathbf{k}}^{e,X} = k^2/2m_{e,X}$ are the electron and exciton dispersions, $m_e$ and $m_X$ are the corresponding effective masses, and $f_{\mathbf{k}}$ is the electron Fermi-Dirac distribution. We use $G_X(\omega)$ to calculate the spectral function $A(\omega) = - \Im G_X(\omega)/\pi$, while the PL spectrum follows from the detailed-balance relation~\cite{kennard1918,tiene2023} $P(\omega) \propto e^{-\hbar\omega/(k_BT)} A(\omega)$. To account for residual broadening not captured by the polaron model, $P(\omega)$ is then convolved with a Lorentzian with full width at half maximum
of $10$~meV,
fixed by the low-temperature experimental linewidth (Supplementary Note~7).

\section{Data availability}

The data that support the findings of this study are openly available in e-cienciaDatos at \url{https://doi.org/XX.XXXXX/YYYYYY}.

\bibliographystyle{unsrtnat}
\bibliography{referencesSiN}

\section{Acknowledgements}

C.P., L.V., M.D.M., F.M.M., and C.A.-S. acknowledge the support from the projects from the Spanish Ministry of Science, Innovation and Universities through the ``Maria de Maeztu'' Programme for Units of Excellence in R\&D (CEX2023-001316-M), projects PID2023-148061NB-I00, PCI2024-153425, and MCIN/AEI/10.13039/501100011033, FEDER UE, project No.~PID2023-150420NB-C31 (Q). K.W. and T.T. acknowledge support from the CREST (JPMJCR24A5), JST and World Premier International Research Center Initiative (WPI), MEXT, Japan. C.A.-S. acknowledges the support from the Comunidad de Madrid fund ``Atracci\'on de Talento, Mod. 1'', Ref. 2020-T1/IND-19785 and the project from the Ministerio de Ciencia e Innovaci\'on CNS2025-165107. M.M.P. is supported through Australian Research Council Future Fellowship FT200100619, and J.L. through Australian Research Council Discovery Project DP240100569. C.P. and C.A.-S. acknowledge the technical support of Fabiola Mogoll\'on and Antonio Flores and the General Support Services for Experimental Research of the UAM (SEGAINVEX).

\section{Author contributions}

A.C.-G.\ proposed the study of SiN$_x$ capping and supervised the sample assembly. R.L.M.\ assembled all the samples. K.W.\ and T.T.\ provided the hBN crystals. C.P.\ performed the experiments, analysed the data, and prepared the figures. A.C.L.\ assisted with the experiments, analysed the data, and prepared the figures. S.P.V.\ assisted with the experiments and the data analysis. B.C.M.\ developed the theory, performed the simulations, and helped prepare the figures. J.L., M.M.P., and F.M.M.\ supervised the theory development, and F.M.M.\ provided theoretical feedback during the experiments. M.D.M.\ co-supervised C.P. C.A.-S., F.M.M., M.M.P. and J.L.\ co-wrote the manuscript, with contributions from all authors. C.A.-S.\ supervised the whole project. All authors discussed the results.

\section{Competing interests}

The authors declare no competing interests.

\clearpage
\setcounter{figure}{0}
\setcounter{table}{0}
\setcounter{equation}{0}
\setcounter{section}{0}
\setcounter{secnumdepth}{2}
\renewcommand{\figurename}{Supplementary Fig.}
\renewcommand{\tablename}{Supplementary Table}
\renewcommand{\theequation}{S\arabic{equation}}
\renewcommand{\thesection}{\arabic{section}}
\renewcommand{\thesubsection}{\arabic{section}.\arabic{subsection}}
\makeatletter
\def\@seccntformat#1{\csname prefix@#1\endcsname\csname the#1\endcsname\quad}
\def\prefix@section{Supplementary Note~}
\def\prefix@subsection{}
\makeatother

\begin{center}
{\Large\bfseries Supplementary Information}
\end{center}

\section{Experimental setups}
\label{sec:setups}

\begin{figure}[htbp]
\centering
\includegraphics[width=1\textwidth]{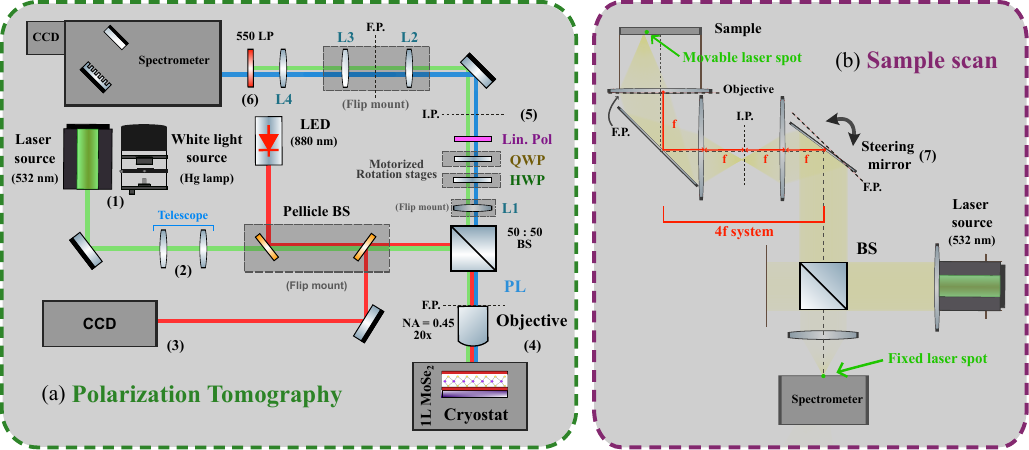}
\caption{\textbf{Optical setups.}
\textbf{a}~Polarization-tomography configuration. (1)~$532$~nm continuous-wave
excitation laser; (2)~beam-expanding telescope; (3)~CCD used for white-light
imaging of the sample through a flip-mounted pellicle beam splitter fed by an
$880$~nm light-emitting diode; (4)~$20\times$, ${\rm NA}=0.45$ objective focusing
onto the 1L~MoSe$_2$ heterostructure held in the cryostat; (5)~polarization-analysis
stage comprising motorized half-wave (HWP) and quarter-wave (QWP) plates followed
by a fixed linear polarizer, with the flip-mounted lens $L_1$ and the relay
$L_2$--$L_3$ selecting between real-space and Fourier-space imaging;
(6)~$550$~nm long-pass filter and spectrometer. F.P.\ and I.P.\ denote Fourier
and image planes, respectively.
\textbf{b}~Sample-scan configuration. A $4f$ relay images the steering mirror~(7) onto
the back focal plane of the objective, so that the excitation spot is translated
across the sample while the collection spot on the spectrometer slit stays fixed.}
\label{figS:setups}
\end{figure}

Supplementary Fig.~\ref{figS:setups} shows the two optical configurations used in this work, described in the Methods of the main text. Both share the same open-loop helium cryostat and detection system. Configuration \textbf{a} is used for the PL and reflectance spectra and, with the lens $L_1$ inserted, for the momentum-resolved polarization tomography of Supplementary Note~\ref{sec:stokes}. Since the open-loop cryostat leaves no space for cryogenic nanopositioners, the sample-scan configuration \textbf{b} is used for the spatially resolved maps of Supplementary Note~\ref{sec:plmaps}: a $4f$ relay images the steering mirror onto the back focal plane of the objective, so that the excitation spot is translated across the sample while the collection path remains fixed~\cite{Lee2007}.

\section{Spatially resolved PL of the SiN$_x$-\\encapsulated sample}
\label{sec:plmaps}

To establish whether the SiN$_x$-induced doping inferred in the main text is spatially uniform, we mapped the low-temperature PL over the region enclosed by the green dashed box in Supplementary Fig.~\ref{figS:plmaps}a, which contains several disconnected 1L~MoSe$_2$ islands within the same SiN$_x$/MoSe$_2$/SiN$_x$ stack. At each point of the map, a linear spectral background was subtracted using the average intensities on the low- and high-wavelength sides of the measured spectral window. The dominant emission energy was obtained from the intensity-weighted average of the spectral points around the emission maximum. The FWHM was extracted from the two half-maximum crossings using linear interpolation between two adjacent spectral points. Lastly, the integrated PL intensity was obtained by numerical integration of the background-subtracted spectrum. The resulting integrated intensity, peak energy, and FWHM are shown in Supplementary Fig.~\ref{figS:plmaps}b, c, and~d, respectively.

The islands are readily identified in the intensity map (Supplementary Fig.~\ref{figS:plmaps}b). The peak energy (Supplementary Fig.~\ref{figS:plmaps}c), lies within $1.625$--$1.632$~eV everywhere, i.e.\ the emission is the attractive-polaron line $\mathrm{AP}$ identified in Fig.~1d of the main text over the entire mapped area; the neutral exciton is nowhere the dominant feature. Although a small spatial variation of the emission energy is observed across and between the individual MoSe$_2$ islands, no region exhibits a shift towards the neutral-exciton energy. This demonstrates that the predominance of the AP emission is not restricted to a particular position on the sample, but persists throughout the spatially separated monolayer regions within the same SiN$_x$/MoSe$_2$/SiN$_x$ stack.

The linewidth map (Supplementary Fig.~\ref{figS:plmaps}d), likewise shows that the AP emission remains with a similar FWHM throughout the mapped monolayer regions. Together, the peak-energy and linewidth maps demonstrate that the spectral characteristics assigned to the SiN$_x$-induced electron doping in the main text are reproduced across multiple spatially disconnected MoSe$_2$ islands, rather than arising from a single locally doped region of the sample.

\begin{figure}[htbp]
\centering
\includegraphics[width=0.7\textwidth]{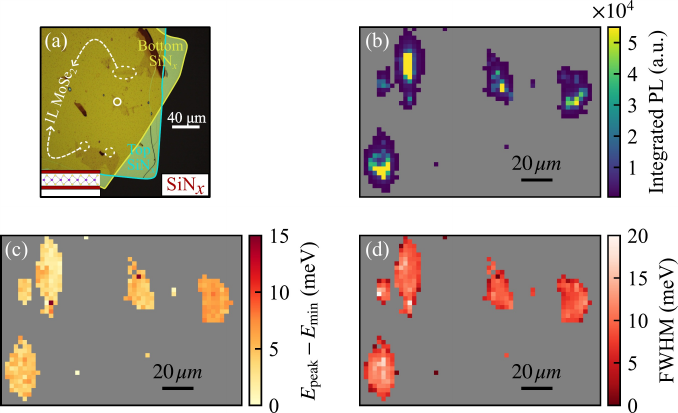}
\caption{\textbf{PL maps of the SiN$_x$-encapsulated sample at $T=8$~K.} \textbf{a}~Optical micrograph of the SiN$_x$/1L~MoSe$_2$/SiN$_x$ stack; the green rectangle marks the region mapped in \textbf{b}--\textbf{d}. The white circle corresponds to the monolayer region used for the PL and reflectance experiments. Scale bar, $40~\mu$m. Inset: schematic cross section of the heterostructure. \textbf{b}~Spectrally integrated PL intensity, revealing the spatially separated 1L~MoSe$_2$ islands within the mapped region. The displayed intensity range was chosen to provide optimal contrast across the different PL-emitting regions; consequently, a small number of high-intensity pixels exceed the upper limit of the colour scale. \textbf{c}~Spatial variation of the peak energy of the dominant emission line, shown as $E_{\mathrm{peak}}-E_{\min}$ in meV, where $E_{\min}=1.620~\mathrm{eV}$ is the minimum energy of the range of interest. The emission remains within the energy range associated with the attractive-polaron line $\mathrm{AP}$ across the different MoSe$_2$ islands. \textbf{d}~Full width at half maximum (FWHM) of the dominant emission line, showing the spatial variation of the AP linewidth across the individual MoSe$_2$ islands. The PL maps were acquired using a $20\times$ Olympus objective with a spatial step size of $1~\mu$m and an estimated excitation power of $0.140$~mW at the sample plane. Grey pixels in \textbf{b}--\textbf{d} denote points where the PL intensity was below the threshold used for reliable extraction of all the spectral parameters. Isolated pixels observed away from the PL-emitting MoSe$_2$ islands, consistently appearing in \textbf{b}--\textbf{d}, are attributed to cosmic-ray events detected by the CCD and do not represent sample emission. Scale bars, $20~\mu$m.}
\label{figS:plmaps}
\end{figure}

\section{Excitation-power dependence}
\label{sec:pdep}

\begin{figure}[htbp]
\centering
\includegraphics[width=1\textwidth]{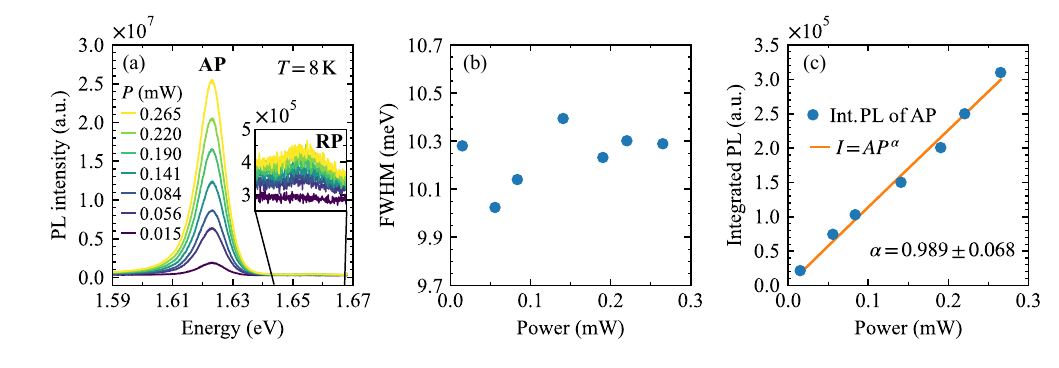}
\caption{\textbf{Excitation-power dependence of the ${\mathrm{SiN}_x}$-encapsulated sample at $T=8$~K.} \textbf{a}~PL spectra for excitation powers between $0.015$ and $0.265$~mW. The spectra are dominated by the attractive-polaron line $\mathrm{AP}$. Inset: magnified view ($\times 10^5$ ordinate scale) of the high-energy region containing the weak repulsive-polaron emission $\mathrm{RP}$. \textbf{b}~Full width at half maximum of the $\mathrm{AP}$ line versus excitation power. \textbf{c}~Integrated PL intensity of the $\mathrm{AP}$ line versus excitation power, integrated over the energy range $1.590$--$1.644$~eV to exclude the higher-energy $\mathrm{RP}$ feature. The solid line shows a fit to a power law, yielding $\alpha=0.989\pm0.068$. Error bars in \textbf{b} and \textbf{c} are smaller than the corresponding data markers and are therefore omitted.}
\label{figS:pdep}
\end{figure}

Supplementary Fig.~\ref{figS:pdep} shows the dependence of the ${\mathrm{SiN}_x}$-encapsulated sample emission on excitation power at $T=8$~K, over the range $0.015$--$0.265$~mW. The spot size presents a diameter of around 2 $\mu$m. Two points relevant to the main text follow from it:

First, the inset of Supplementary Fig.~\ref{figS:pdep}a resolves the weak $\mathrm{RP}$ emission near $1.655$~eV that is at the limit of detection in Fig.~1d of the main text. Its integrated intensity is smaller than that of $\mathrm{AP}$ by roughly two orders of magnitude at every power investigated, and the ratio does not change appreciably over the range studied. The inversion of spectral weight is therefore an intrinsic property of the doped monolayer and not an artifact of the excitation conditions.

Second, the emission linewidth of $\mathrm{AP}$ (Supplementary Fig.~\ref{figS:pdep}b), is constant within $\pm0.2$~meV over an eighteen-fold change in excitation power, while the integrated intensity (Supplementary Fig.~\ref{figS:pdep}c), grows monotonically. The measurements reported in the main text are thus taken in a regime where neither excitation-induced heating nor photo-induced screening of the resident carriers modifies the line shape, so that the temperature-dependent linewidths of Supplementary Fig.~\ref{figS:Tdep}f can be attributed to the intrinsic temperature dependence.

To quantify the excitation-power dependence, the integrated intensity of the $\mathrm{AP}$ emission was evaluated over the energy interval $1.590$--$1.644$~eV. The upper integration bound was chosen to isolate the $\mathrm{AP}$ branch and exclude the weak higher-energy $\mathrm{RP}$ feature. A constant baseline defined by the minimum measured intensity of each spectrum was used for the integration. No additional background subtraction was applied, since the remaining background could not be reliably determined without substantially affecting the very weak higher-energy emission. The resulting integrated intensity was fitted with a power law yielding $\alpha=0.989\pm0.068$, consistent with a linear increase of the $\mathrm{AP}$ emission with excitation power throughout the investigated range. The uncertainties associated with both the extracted FWHM and the integrated PL intensity are smaller than the corresponding data markers and are therefore omitted from Supplementary Fig.~\ref{figS:pdep}b,c for clarity. Together, the approximately linear power dependence and nearly power-independent linewidth show no evidence of pronounced saturation or excitation-induced modification of the $\mathrm{AP}$ emission over the range of powers investigated.

\section{Temperature dependence of the SiN$_x$-encapsulated sample}
\label{sec:Tdep}

\begin{figure}[htbp]
\centering
\includegraphics[width=0.9\textwidth]{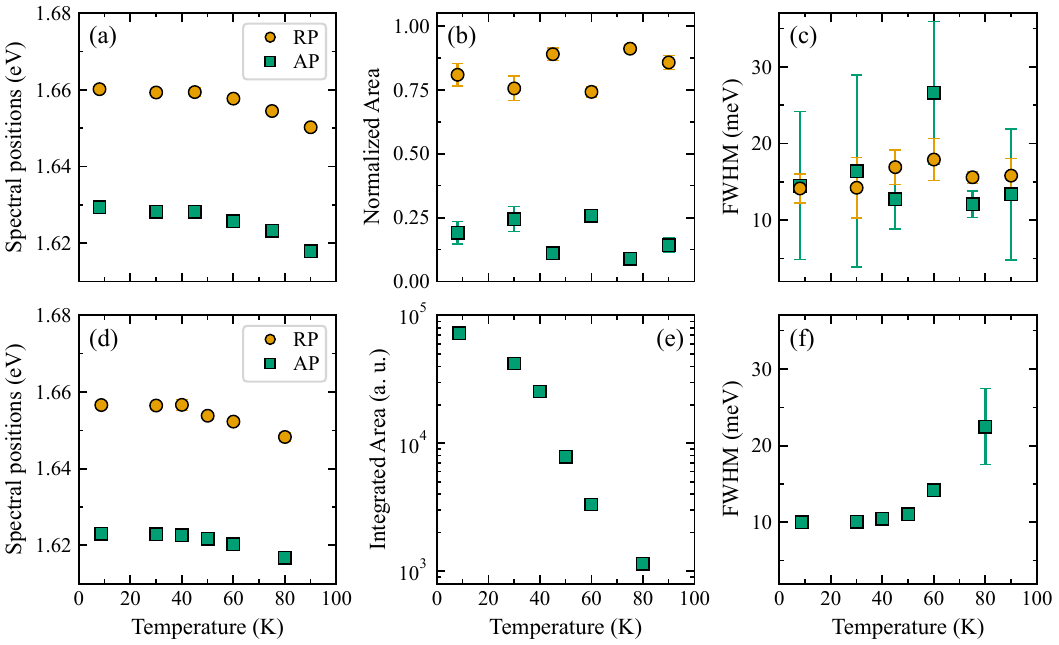}
\caption{\textbf{Temperature dependence of $\Delta R/R$ and PL of the SiN$_x$-encapsulated sample.} Top row, \textbf{a}--\textbf{c}: $\Delta R/R$ analysis; bottom row, \textbf{d}--\textbf{f}: PL analysis. Orange circles, repulsive polaron (RP); blue squares, attractive polaron (AP). \textbf{a}~Energies of the reflectivity minima. \textbf{b}~Normalized area of the two resonances (sum to unity). \textbf{c}~FWHM of the reflectivity resonances. \textbf{d}~PL peak energies. \textbf{e}~Integrated PL intensity of the AP line (log scale). \textbf{f}~FWHM of the AP PL line. Error bars are shown where larger than the markers.}
\label{figS:Tdep}
\end{figure}

Supplementary Fig.~\ref{figS:Tdep} shows the spectral parameters extracted from the reflectivity and PL spectra of Figs.~2a and~3 of the main text, which are not repeated here.

Energies are those of the dip minima in reflectivity (Supplementary Fig.~\ref{figS:Tdep}a) and of the peak maxima in PL (Supplementary Fig.~\ref{figS:Tdep}d). The FWHM (Supplementary Fig.~\ref{figS:Tdep}c,f), is the full width at half maximum above the local baseline. The reflectivity areas (Supplementary Fig.~\ref{figS:Tdep}b), are normalized to sum to unity, which removes the common optical prefactor of the stack, and the PL intensity (Supplementary Fig.~\ref{figS:Tdep}e), is the area of the AP line over a window excluding the RP. Two limitations apply. The dips are slightly asymmetric because of a dispersive admixture from the stack, which gives the absolute energies a systematic uncertainty of a few meV that should largely cancel in the temperature dependence and in the AP--RP splitting. The widths of the weak AP dip and of the $80$~K PL line, which overlaps with a low-energy tail and the RP emission, are poorly constrained, as the error bars show. The PL intensity is comparable between temperatures only if the collection efficiency stays constant. The error bars of the AP--RP splitting and of the normalized areas (Fig.~2b,c of the main text), reflect the uncertainty associated with the choice of baseline and of the integration limits of the AP and RP resonances; they were estimated by the Monte Carlo procedure described in the Methods of the main text.

In reflectivity both resonances redshift in parallel (Supplementary Fig.~\ref{figS:Tdep}a), from $1.660$ to $1.650$~eV (RP) and from $1.629$ to $1.618$~eV (AP) between $8$ and $90$~K, accelerating above $\sim45$~K as expected for a semiconductor band gap~\cite{vina1984}, so that the splitting stays at $\approx31$--$32$~meV (Fig.~2c of the main text). The RP keeps $\approx75$--$90\%$ of the weight with no discernible trend, Supplementary Fig.~\ref{figS:Tdep}b; a thermal release of carriers into the nitride would instead transfer weight from the AP to the RP, so the persistent AP weight is consistent with a constant density. The RP FWHM is $\approx14$--$18$~meV without a clear trend, whereas the AP FWHM ($\approx13$--$27$~meV, errors up to $\sim\pm20$~meV) is too uncertain to show one (Supplementary Fig.~\ref{figS:Tdep}c).

In PL the AP and RP lines redshift~\cite{vina1984} (Supplementary Fig.~\ref{figS:Tdep}d), from $1.623$ to $1.617$~eV and from $1.657$ to $1.648$~eV between $8.7$ and $80$~K, and lie $\approx6$ and $\approx3$~meV below the reflectivity minima at low temperature. The AP intensity (Supplementary Fig.~\ref{figS:Tdep}e), falls by a factor of $\sim60$, slowly below $\sim30$~K and steeply above, with an apparent activation energy of $\sim20$~meV between $50$ and $80$~K that we do not assign to a specific channel. The RP emission grows relative to the AP, as the magnification needed to display it in Fig.~3 of the main text falls from $\times180$ at $8.7$~K to $\times7$ at $60$~K. The AP FWHM (Supplementary Fig.~\ref{figS:Tdep}f), stays at $10$--$11$~meV up to $50$~K and then grows to $14$ and $22$~meV at $60$ and $80$~K, the latter less certain. This onset coincides with the temperature, $T\approx55$--$60$~K, at which the crossover condition $E_F\approx0.7\,k_BT$ is met for the extracted density ($E_F\approx3.2$--$3.6$~meV), and is consistent with the crossover to the trion-dominated regime discussed in the main text. Altogether, the AP persists in absorption up to $90$~K with an almost constant weight and splitting from the RP, while the emission is quenched and broadened thermally.

\section{Dielectric environment and range of trion binding energies}
\label{sec:dielectric}

The trion binding energy $\varepsilon_T$ enters the Fermi-polaron model as an input parameter, and no calculation of it is available for MoSe$_2$ encapsulated in amorphous silicon nitride. This section summarizes the arguments that restrict the range explored in the main text, $\varepsilon_T=24$--$28$~meV.

Each membrane is $50\pm5$~nm thick~\cite{tedpella}, about $45$ times the r.m.s.\ radius of the MoSe$_2$ exciton, $\approx1.1$~nm~\cite{goryca2019}, and well above the $\lesssim10$~nm range in which the screening of the nitride is reduced near its surfaces~\cite{pham2011}. Each membrane therefore screens the monolayer as a bulk dielectric, characterized by its optical dielectric constant $\epsilon_\infty\approx n_r^2$, with $n_r$ the refractive index below the gap. The supplier considers the deposition method and stoichiometry of the films proprietary and has no data on their refractive index; $n_r$ must therefore be bracketed with literature values, which depend on stoichiometry and deposition method. At the exciton wavelength, $\approx750$~nm, stoichiometric CVD and LPCVD Si$_3$N$_4$ give $n_r\approx2.00$--$2.03$~\cite{philipp1973,luke2015}, i.e.\ $\epsilon_\infty\approx4.0$--$4.1$, and a $100$-nm PECVD film gives $n_r\approx2.01$~\cite{beliaev2022}, whereas Si-rich, low-stress LPCVD nitride, the type commonly used for membranes, reaches $\epsilon_\infty\approx4.5$ in $100$-nm films~\cite{beliaev2022}. The supplier describes its ultra-low-stress films, which include our $50$-nm membranes, as non-stoichiometric and closer to SiN than to Si$_3$N$_4$~\cite{tedpella}, so we expect $\epsilon_\infty\approx4.0$--$4.6$ for our nitride, similar to the value of hBN, $\epsilon_\infty\approx4.4$--$4.5$~\cite{goryca2019}.

Replacing hBN with SiN$_x$ should therefore change $\varepsilon_T$ only slightly. For comparison, the trion binding energy of MoSe$_2$ changes only from $\approx30$~meV on SiO$_2$~\cite{ross2013} to $\approx25$~meV in our hBN reference (Fig.~1d of the main text), although the screening changes much more between these two environments. Since the X$^0$--X$^-$ separation measured in PL includes the doping effects of residual electrons, it is an upper bound for the trion binding energy in hBN~\cite{mak2013}. The range $\varepsilon_T=24$--$28$~meV, centered close to this value, therefore covers the expected binding energy in the nitride. The value extracted from the AP--RP splitting, $\varepsilon_T\approx26$~meV, lies in the middle of this range.

\section{Momentum-resolved polarization tomography}
\label{sec:stokes}

Encapsulation is not polarization-neutral. In angle-resolved measurements on MoSe$_2$ monolayers, hBN encapsulation was found to imprint a pronounced linear polarization on the exciton emission at large emission angles, reaching a degree of linear polarization of $\approx 17\%$, an effect traced by finite-difference time-domain modelling to the uniaxial optical anisotropy of the hBN layers themselves; in a suspended monolayer, where that extrinsic contribution is absent, the residual momentum-dependent polarization is several times smaller and appears only at cryogenic temperatures, where it is attributed to the intrinsic $k$-dependent electron--hole exchange coupling that splits the exciton dispersion into TE and TM branches \cite{Han2022}. Amorphous SiN$_x$ has no crystalline optic axis, so the hBN-like extrinsic contribution should be absent here; conversely, residual stress in the membranes could in principle imprint a preferred in-plane axis. Both possibilities are tested by the same measurement.

We performed a full Stokes tomography of the $\mathrm{AP}$ emission in the Fourier plane, using the configuration of Supplementary Fig.~\ref{figS:setups}a with the lens $L_1$ inserted. The full $\mathrm{AP}$ line is spectrally integrated at each point, so the analysis is sensitive to the polarization texture in momentum space but not to any underlying energy splitting of the emitting branches. The normalized Stokes parameters $S_1$, $S_2$ and $S_3$, reconstructed pixel by pixel from six polarization projections, are shown as functions of in-plane momentum in Supplementary Fig.~\ref{figS:stokes}a--c, and the degree of polarization ${\rm DOP} = \sqrt{S_1^2+S_2^2+S_3^2}$ in Supplementary Fig.~\ref{figS:stokes}d.

All three Stokes parameters remain below $0.15$ in magnitude everywhere within the collection numerical aperture, and the circular component $S_3$ (Supplementary Fig.~\ref{figS:stokes}c), is featureless at the noise level, consistent with the non-resonant, above-gap excitation used here, which does not preserve valley polarization. Azimuthally binning the maps (Supplementary Fig.~\ref{figS:stokes}e), reveals a weak but systematic modulation of $S_1$ with a period of $\pi$ in the in-plane angle $\phi$, of amplitude $\approx 0.05$, superposed on a degree of polarization that is independent of $\phi$ and averages $0.066$.

A fixed in-plane axis, such as the one that would be imprinted by uniaxial residual stress in the membranes, produces a Stokes vector that is constant in the laboratory frame and therefore a $\phi$-\emph{independent} $S_1$. The $\pi$-periodic $S_1(\phi)$ observed here, accompanied by a $\phi$-independent DOP, is instead the signature of a polarization axis that rotates with the azimuthal angle, i.e.\ of an imbalance between the $p$- and $s$-polarized (equivalently TM- and TE-like) components of the far field, which is precisely the decomposition used in Ref.~\cite{Han2022}. The residual anisotropy is thus of the same nature as that reported for suspended MoSe$_2$ monolayers, and not a uniaxial strain axis.

The amplitude found here, $\approx 0.05$, is roughly three times smaller than the $\approx 17\%$ reported for hBN-encapsulated MoSe$_2$ at the edge of a comparable collection cone, and is instead of the order of the suspended case \cite{Han2022}. This is what the absence of a crystalline optic axis in amorphous SiN$_x$ would predict: the nitride does not add the extrinsic, birefringence-driven polarization anisotropy that hBN does. In this regard, SiN$_x$ encapsulation is the \emph{less} perturbative of the two environments, which is a favourable property for the polarization- and valley-resolved experiments that an integrated platform is ultimately meant to support.

Finally, the same Fourier-plane data bound the orientation of the emitting dipole. An out-of-plane dipole component radiates no power along the surface normal and contributes purely radially polarized light that grows steeply towards the edge of the collection cone, so that the in- and out-of-plane dipole weights can be extracted from the momentum dependence of the polarized far field \cite{BrotonsGisbert2019,Sigl2022}. We observe no such steep radial component within our numerical aperture, consistent with the in-plane dipole orientation established for monolayer MoSe$_2$ \cite{BrotonsGisbert2019} and for MoSe$_2$--WSe$_2$ interlayer excitons, which are $(99\pm1)\%$ in plane \cite{Sigl2022}. The attractive polaron therefore radiates as an in-plane dipole, like the neutral exciton, and contact with the nitride does not introduce out-of-plane dipole character.

\begin{figure}[htbp]
\centering
\includegraphics[width=\textwidth]{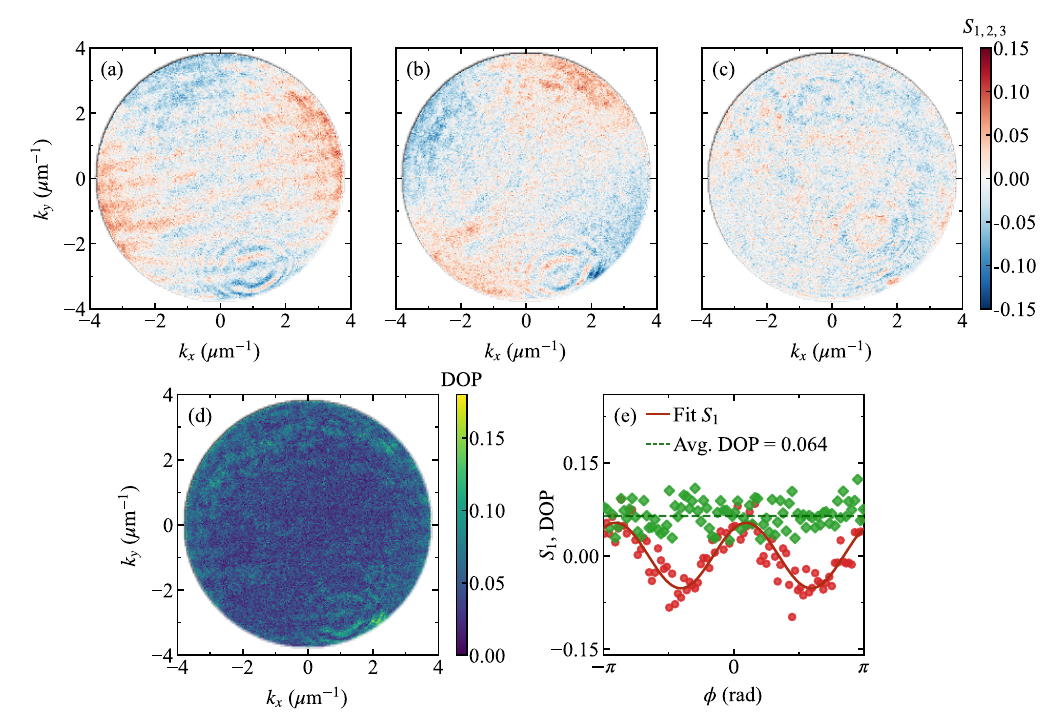}
\caption{\textbf{Momentum-resolved Stokes tomography of the SiN$_x$-encapsulated sample.} \textbf{a}--\textbf{c}~Normalized Stokes parameters $S_1$, $S_2$ and $S_3$ of the $\mathrm{AP}$ emission as functions of the in-plane momentum $(k_x,k_y)$. The $S_3$ parameter showed no significant polarization features, indicating the absence of a pronounced circular-polarization component. \textbf{d}~Corresponding degree of polarization, ${\rm DOP}=\sqrt{S_1^2+S_2^2+S_3^2}$. The ring-shaped feature observed near $(k_x,k_y) \approx (1,-3)~\mu\mathrm{m}^{-1}$ in the maps \textbf{a}--\textbf{d} is attributed to an experimental artifact, possibly arising from a dust particle or an etalon fringe in the detection path. \textbf{e}~Azimuthal dependence of $S_1$ (red circles) and of the DOP (green diamonds), obtained by binning the maps in \textbf{a} and \textbf{d} as a function of the in-plane angle $\phi$. The red dash-dotted line is a fit to $S_1$ with a $\pi$-periodic modulation; the green dashed line marks the angle-averaged degree of polarization, $0.064$. The Stokes analysis was performed at $10$~K with an excitation power of $0.215$~mW at the sample plane. The $k$-space scale was determined considering PL emission centered at $759$~nm and an objective with a numerical aperture of $\mathrm{NA}=0.45$. The corresponding maximum in-plane momentum was estimated to be $k_{\parallel,\max} \approx 3.73~\mu\mathrm{m}^{-1}$.}
\label{figS:stokes}
\end{figure}

\subsection{Azimuthal analysis of the Stokes parameters $S_1$ and $S_2$}
\label{sec:azimuthal_stokes}

To investigate the momentum dependence of the polarization texture, the $S_1$ and $S_2$ maps were analyzed along annuli of fixed $|\mathbf{k}|$. Each annulus corresponds to a constant emission angle $\theta$, with its radius in the Fourier-plane image determined according to $r(\theta)=R_{\mathrm{NA}}\sin\theta/\mathrm{NA}$, where $R_{\mathrm{NA}}$ is the radius corresponding to the numerical aperture of the objective ($\mathrm{NA}=0.45$). Eight equally spaced emission angles were considered, ranging from $\theta=15^\circ$ to the largest angle for which the complete annulus remained within the objective NA. An annular half-width of two pixels was used for each emission angle, i.e.\ $\pm2$ pixels around the central radius. Within each annulus, the Stokes parameter was evaluated as a function of the in-plane azimuthal angle $\phi$ in momentum space.

The data were grouped into $2^\circ$ azimuthal bins, and the mean Stokes parameter within each bin was calculated as $\overline{S_i}(\phi)
=\frac{1}{N}\sum_{j=1}^{N} S_{i,j}$, where $S_{i,j}$ is the Stokes parameter of the $j$th pixel within the corresponding azimuthal bin and $N$ is the number of contributing pixels. Only bins containing at least three valid pixels were retained. We take the standard deviation $\sigma$ of the Stokes parameter to calculate the uncertainty of the mean Stokes parameter of each bin as the standard error of the mean (SEM) $\sigma/\sqrt{N}$.

For each emission angle, the azimuthal dependence was first fitted directly to the binned mean Stokes parameters using the twofold angular modulation $S_i(\phi)=A\cos\left[2(\phi-\phi_0)\right]+C$, $i=1,2$, where $A$ is the modulation amplitude, $\phi_0$ is the phase of the angular modulation, and $C$ is a constant offset. In this initial fit, $A$, $\phi_0$, and $C$ were treated as free parameters. When valid SEM values were available for at least four azimuthal bins, the fit was weighted by the corresponding SEM values; otherwise, an unweighted fit was performed. This initial fitting step therefore determined the best-fit parameters $A_{\mathrm{best}}$, $\phi_{0,\mathrm{best}}$, and $C_{\mathrm{best}}$ directly from the azimuthally-binned mean Stokes parameters and their associated SEM values. The modulation amplitude was defined to be positive, with the corresponding phase adjusted accordingly.

The asymmetric uncertainties in $A_{\mathrm{best}}$ were determined in a subsequent step using the SEM of the azimuthally binned data. First, the binned Stokes values were separated into positive and negative lobes according to whether they lay above or below the best-fit offset $C_{\mathrm{best}}$. An outer SEM envelope was then constructed by adding the SEM to the positive-lobe values and subtracting the SEM from the negative-lobe values, thereby increasing their excursions from $C_{\mathrm{best}}$. Conversely, an inner SEM envelope was constructed by subtracting the SEM from the positive-lobe values and adding the SEM to the negative-lobe values, thereby reducing their excursions from $C_{\mathrm{best}}$.

The outer and inner SEM envelopes were subsequently fitted independently while fixing $\phi_0=\phi_{0,\mathrm{best}}$ and $C=C_{\mathrm{best}}$, such that only the modulation amplitude $A$ was allowed to vary. The fit to the outer envelope yielded $A_{\max}$, whereas the fit to the inner envelope yielded $A_{\min}$. The asymmetric uncertainties of the best-fit modulation amplitude were then defined as $\Delta A_{-}=A_{\mathrm{best}}-A_{\min}$ and $\Delta A_{+}=A_{\max}-A_{\mathrm{best}}$. The resulting best-fit modulation amplitudes $A_{\mathrm{best}}$ and their asymmetric uncertainties were finally plotted as a function of the emission angle $\theta$ to examine the evolution of the polarization modulation with increasing emission angle and, equivalently, increasing in-plane momentum.

\begin{figure}[htbp]
\centering
\includegraphics[width=1\textwidth]{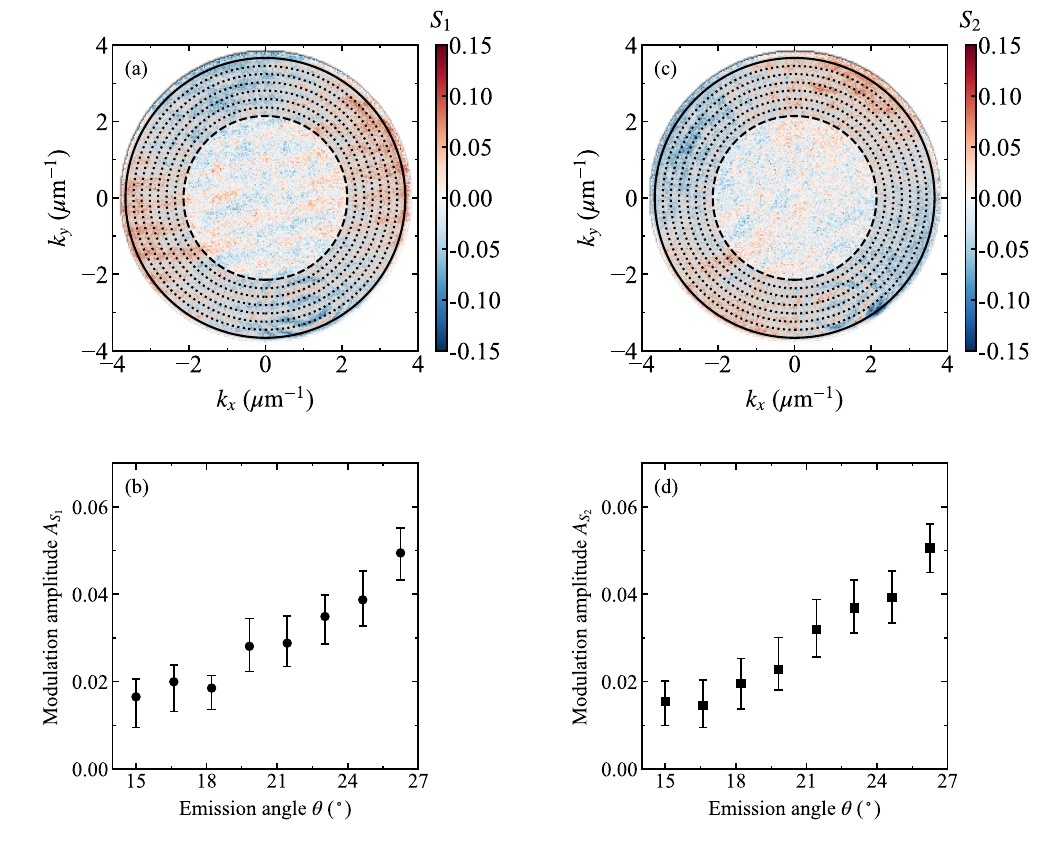}
\caption{\textbf{Momentum-resolved azimuthal analysis of the Stokes parameters $\mathbf{S_1}$ and $\mathbf{S_2}$.} \textbf{a} shows the $S_1$ map together with the fixed-$|k|$ annuli used for the analysis, spanning increasing emission angles from the inner (dashed line) to the outer region (solid line) of the accessible $k$-space. \textbf{b} shows the corresponding modulation amplitude $A$ (dark circles) extracted from the azimuthal fits as a function of emission angle $\theta$, with asymmetric error bars determined from the SEM-envelope analysis. \textbf{c} and \textbf{d} show the equivalent analysis for $S_2$, where the modulation amplitude is illustrated in dark squares.}
\label{figS:Azimuthal_stokes}
\end{figure}

The amplitude of the azimuthal modulation increases with emission angle and, consequently, with the in-plane momentum $k_{\parallel}=(2\pi/\lambda)\sin\theta$. In particular, the fitted $S_1$ modulation amplitude increases from $1.65\%$ at $\theta=15^\circ$ to $4.95\%$ at $\theta=26.25^\circ$, corresponding to an approximately threefold increase over the investigated momentum range. At the largest emission angle, the $S_2$ modulation reaches a comparable value of $5.06\%$, independently confirming the $\sim5\%$ maximum linear-polarization modulation obtained from $S_1$. Furthermore, the fitted phases of $S_1$ and $S_2$ differ by approximately $42^\circ$ on average, close to the $45^\circ$ phase difference expected between the two Stokes components of a rotating linear-polarization texture.

The amplitude of the $S_1$ modulation can be directly compared with the $s/p$-resolved degree of linear polarization (DOLP) reported by Han \textit{et al.}~\cite{Han2022}, as both provide a measure of the azimuthally varying linear-polarization contrast in momentum space. Han \textit{et al.} observed a similar momentum dependence in monolayer MoSe$_2$, with the amplitude of the DOLP increasing monotonically towards larger $k_{\parallel}$ \cite{Han2022}. In particular, a DOLP modulation amplitude of approximately $6.7\%$ was reported at $\theta=25^\circ$ for suspended monolayer MoSe$_2$ at $10$~K, in close agreement with the $S_1$ modulation of $4.95\%$ measured here at the comparable emission angle of $\theta=26.25^\circ$. For suspended MoSe$_2$, this momentum-dependent polarization response was associated with the TE--TM exciton splitting arising from long-range exchange interactions. The increase of the $S_1$ modulation with emission angle therefore supports a momentum-dependent origin of the observed linear-polarization texture, with both its magnitude and momentum dependence being comparable to those reported for suspended MoSe$_2$, while remaining substantially smaller than the enhanced response observed for the hBN-encapsulated structure.

\section{Impact of the parameters of the Fermi-polaron model on the PL simulations}
\label{sec:theory}
\begin{figure}[htbp]
\centering
\includegraphics[width=\textwidth]{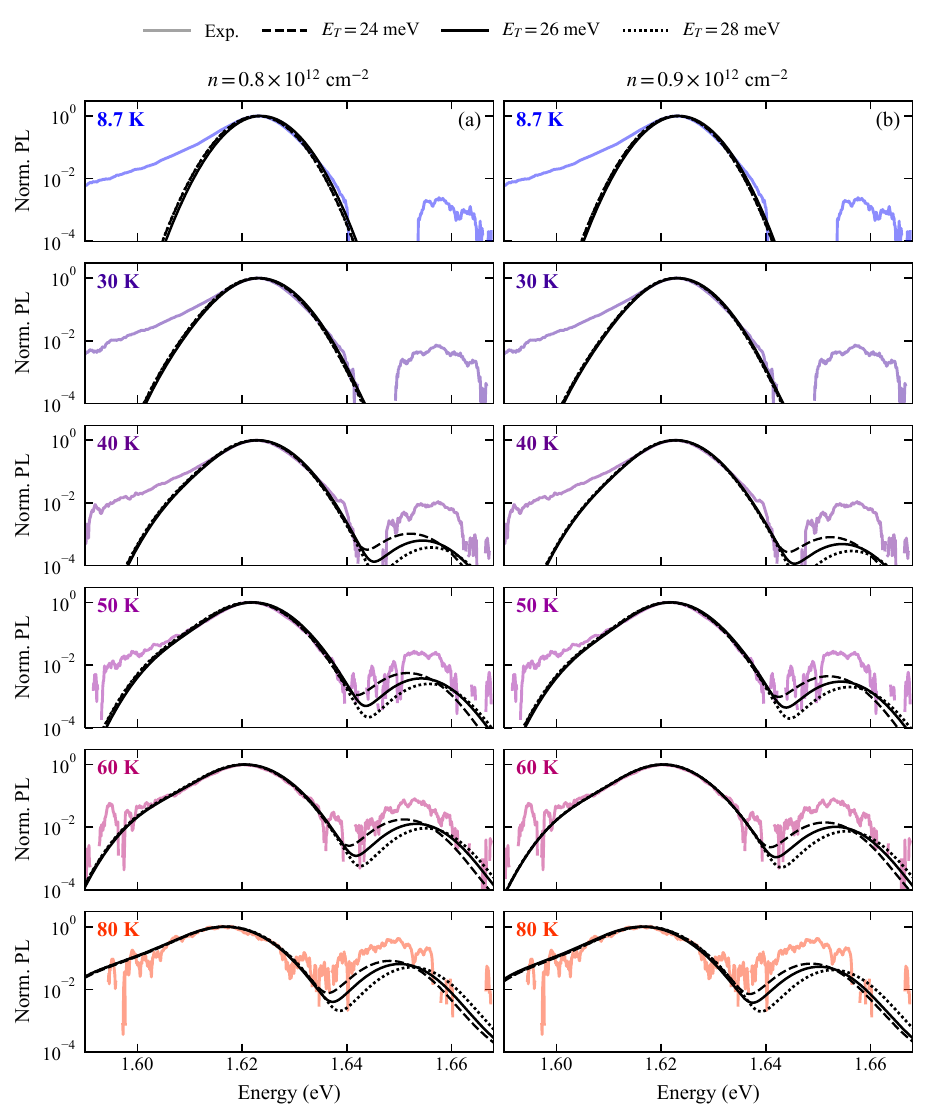}
\caption{Normalized PL spectra on a logarithmic scale from 8.7 to 80~K for electron densities (a) $n = 0.8\times10^{12}$~cm$^{-2}$ and (b) $n = 0.9\times10^{12}$~cm$^{-2}$. Colored lines: experimental spectra after subtraction of a linear background and normalization to the maximum of the AP peak. Black lines: calculated spectra for trion binding energies $E_T = 24$ (dashed), 26 (solid) and 28~meV (dotted), with a broadening of $\eta = 5$~meV. The calculated spectra are shifted in energy at each temperature so that their AP peak coincides with the experimental one. The logarithmic scale reveals the weak RP branch at higher energy, whose position and intensity depend on $E_T$; for $T \leq 30$~K the calculated RP intensity is below $10^{-4}$ and therefore not visible.}
\label{figS:PL-comparison-theory}
\end{figure}

In the main text, we extracted the doping density $n$ and the trion binding energy $\varepsilon_T$ by comparing the temperature dependence of the RP--AP energy splitting and the oscillator strengths $Z_{\mathrm{AP,RP}}$ extracted from reflectance data with the predictions of the Fermi polaron model~\cite{tiene2023}. This analysis allowed us to extract the parameter ranges $n = (0.8$-$0.9)\times 10^{12}$~cm$^{-2}$ and $\varepsilon_T = 24-28$~meV. Here, we show in Fig.~\ref{figS:PL-comparison-theory} that the expected PL spectra, calculated for these ranges of doping density and trion binding energy, exhibit very similar features and are in good agreement with the experimental data.

\end{document}